\documentclass[onecolumn,showpacs,floatfix]{revtex4-1}
\usepackage[utf8]{inputenc}
\usepackage[T1]{fontenc}
\usepackage{amssymb}
\usepackage{amsmath}
\usepackage{graphicx}
\usepackage{color}
\usepackage{bm}
\usepackage{float}
\usepackage[normalem]{ulem}
\usepackage{natbib}

\begin{document}
\title{Oscillations of skyrmion clusters in Co/Pt multilayer nanodots}

\author{Felipe Tejo$^{1}$}\email{felipe.tejo@usach.cl}
\author{Felipe Velozo$^1$}
\author{Ricardo Gabriel Elías$^1$}
\author{Juan Escrig$^{1,2}$}

\affiliation{$^{1}$Departamento de F\'isica, Universidad de Santiago de Chile, 9170124 Santiago, Chile}
\affiliation{$^{2}$Center for the Development of Nanoscience and Nanotechnology, 9170124 Santiago, Chile}

\begin{abstract}
In this work we study the oscillations of the skyrmion cores in a multilayer nanodot as a function of the number of skyrmions hosted in the system. When all the skyrmions in the nanodot have the same core radius, and after applying a perpendicular spin-polarized current, a relaxation process takes place towards an equilibrium configuration that is accompanied by coherent damped oscillations of the skyrmion cores, whose frequency depends on the number of skyrmions present in the nanodot. Additionally, we found that the oscillation frequency is directly related to the total energy of the system.
\end{abstract}

\flushbottom
\maketitle
\thispagestyle{empty}

\section*{Introduction}
Magnetic skyrmions are stable or metastable quasiparticles found in two-dimensional or quasi-two-dimensional magnetic systems \cite{Finocchio_2016}. During the last decades, both the behavior and stability of skyrmions have been intensively studied in a wide range of situations and geometries, where the chiral term of energy known as Dzyaloshinskii-Moriya (DMI) interaction, which comes from spin-orbit interaction in the bulk or in an interface, is essential for their stabilization. This difference in the origin of the DMI is responsible for different skyrmionic configurations: when we are in the presence of the bulk DMI, the so-called Bloch skyrmions appear, where the magnetization has a vortex-like chirality, while for the interfacial DMI the N\'eel skyrmions appear, characterized by a magnetization field that is radial when it is projected onto the sample plane.

It is well known that both the oscillations of quasiparticles and the behavior of fluctuations around them are a complex and rich phenomenon found in different physical contexts and where many factors, such as geometry, topology and the type of simulation, play a role. This is also true in magnetic systems where linear oscillations are known as spin waves \cite{stancil-2009aa}. In fact, spin waves have been investigated around skyrmions \cite{doi:10.1063/1.4933407}, in skyrmion lattices \cite{PhysRevB.84.214433} and at Bloch points \cite{PhysRevB.90.224414}, to name a few examples, giving rise to a subfield that mixes spin waves with spintronics, called magnonics \cite{chumak-2015aa}, whose applications are very promising, those ranging from the detection of magnetic solitons \cite{CARVALHOSANTOS2015364} to the control of magnetic structures for information processing \cite{chumak-2015aa}.

In the same direction, in recent years, different research groups have faced several problems related to the oscillations of skyrmions, such as the breathing modes of individual skyrmions \cite{PhysRevB.99.054430} under the action of an alternating external magnetic field \cite{PhysRevB.90.064410}, as well as oscillations and rotation of arrays of skyrmions (called clusters) under in-plane external magnetic fields \cite{PhysRevB.95.134442}. In all these works the nature of the stimulation, the coupling of the magnons \cite{PhysRevLett.120.237203} with the breathing modes of the skyrmions \cite{Leliaert_2018}, and the geometry and nature of the skyrmions themselves seem to be relevant. Recently it was observed that skyrmions leave their equilibrium position and perform a gyrotropic movement using an oscillating out-of-plane magnetic field \cite{doi:10.1063/1.5006681}. The authors also argue that this movement is a consequence of the change in the diameter of the skyrmion along the spiral path.

\begin{figure}
\centering
\includegraphics[width=.6\textwidth]{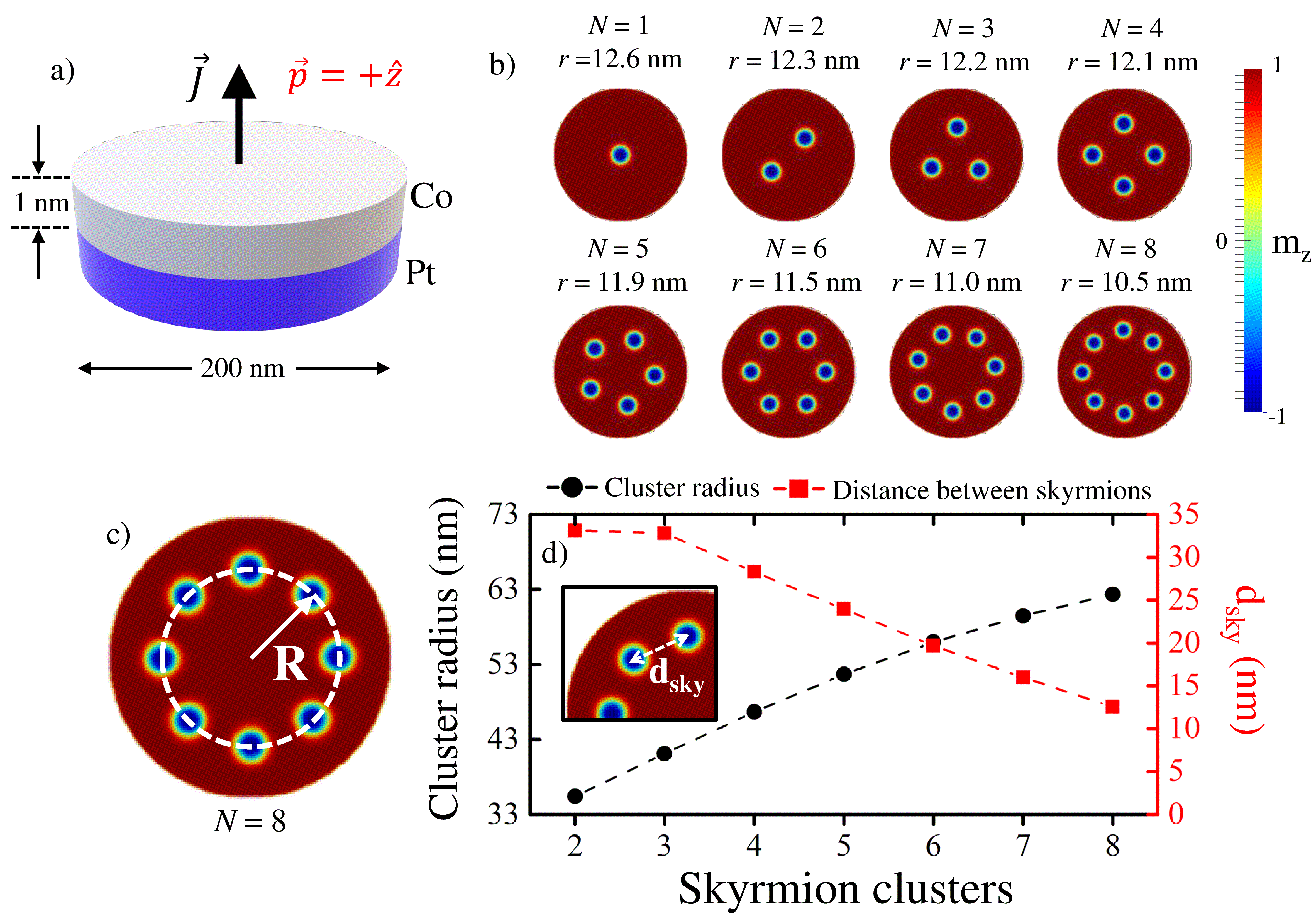} 
\caption{(Color online) a) Schematic representation of the system under study. b) For the parameters used, metastable initial conditions are shown after a relaxation process and before injecting a spin-polarized current. It can be seen that the skyrmion radius $r$ varies little with the number of skyrmions $N$. c) Schematic representation of the definition of the cluster radius. d) Cluster radius ($R$) and distance between skyrmions ($d_{sky}$) as a function of $N$.}
\label{fig1}
\end{figure}

In this paper we are mainly interested in the natural frequency of the first internal breath mode, or uniform breathing mode \cite{PhysRevB.90.064410, Lin2014}, that appears when the core radius of the N\'eel skyrmions changes. The main goal is to explore the influence of the number of skyrmions in a Co/Pt multilayer nanodot \cite{PhysRevLett.99.217208}, which alternates cobalt and platinum layers, on the oscillation frequency of the skyrmion cores. For the process of exciting the multilayer nanodot with an out-of-plane spin-polarized current, the system is modeled by a 1 nm thick cobalt layer \cite{sampaio-2013aa} with the parameters and energy terms given by the actual values of a Co/Pt multilayer nanodot, obtaining in-phase oscillations of the cores (see Fig. \ref{fig1}a). Once we establish the way to obtain coherent oscillations, we numerically investigate the relationship between the number of skyrmions and the frequency of their oscillations. 

This paper is organized as follows: In the second section we present the model and the equations used in the simulations. In the third section we describe the relaxation process that gives rise to the metastable states characterized by the skyrmion cluster. In the fourth section we present and discuss the oscillation process after the current injection. Finally, in the last section we present the conclusions and perspectives of the work.

\section*{Relaxation process of initial configuration}

We consider a Co/Pt multilayer nanodot of 200 nm in diameter, which has a strong interfacial DMI that favors the appearance of N\'eel skyrmions in the 1 nm thick ferromagnetic layer (see Fig. \ref{fig1}a). The ferromagnetic layer is described by the magnetization $\bm M(\bm x,t)=M_s \bm m(\bm x,t)$, where $M_s$ is the saturation magnetization and $|\bm m(\bm x,t)|=1$. The dynamics of this field is governed by the Landau-Lifshitz-Gilbert (LLG) equation \cite{Landau-1935fk}
\begin{equation}
\frac{d\bm m}{dt}= -\gamma\bm{m}\times \bm{H}_{\text{eff}}+ \alpha\bm{m}\times \frac{d\bm{m}}{dt},
\label{eq_LLG}
\end{equation}
where $\gamma$ is the gyromagnetic ratio, $\alpha$ is the Gilbert's damping, responsible of the alignment of the magnetization with the effective magnetic field $\bm{H}_{\text{eff}}$, which can be deduced from the free energy functional, $U[\bm m]$, as \cite{L.-D.-Landau-1984kx}
\begin{equation}
\bm{H}_{\text{eff}}=-\dfrac{1}{\mu_0 M_s}\dfrac{\delta U[\bm m]}{\delta \bm{m}},
\end{equation}
where $\mu_0$ is the vacuum permeability. The energy depends on both the bulk and the interfaces, as well as the geometry and interaction with the external fields. In this study, we have considered the following energy density $\epsilon$.
\begin{eqnarray}
\label{Edensity}
\centering
\epsilon[\bm m]&= A  \sum_{\alpha=x,y,z}(\bm \nabla m_\alpha)^2 +D  [m_z \bm{\nabla}\cdot\bm m-(\bm m \cdot \bm{\nabla})m_{z}]-K_a  m_z^2 -\dfrac{M_s}{2}\mu_{0}\bm{m}\cdot\bm{H}_d-M_s\bm{m}\cdot\bm{B},
\end{eqnarray}
which is related to the functional energy through $U[\bm m]=\int \epsilon [\bm m] dV$. In the definition of Eq. (\ref{Edensity}), the first term corresponds to the exchange energy with the stiffness constant $A$, the second term corresponds to the interfacial DMI, where the magnitude of $D$ controls the strength of the interfacial DMI and its sign determines the chirality of the configurations that minimize its energy. Additionally, we include the effects of out-of-plane uniaxial magnetic anisotropy (where $K_a$ is the uniaxial anisotropy constant), dipolar interaction and Zeeman energy, where $\bm B$ is the Oersted field produced by the current flowing through the sample.

Once the energies involved are defined, we consider the following magnetic parameters for the multilayer Co/Pt system \cite{sampaio-2013aa}: $M_s=580\cdot10^3 \text{(A/m)}$, $A=15\cdot10^{-12}\text{(J/m)}$, $K_a=0.7\cdot10^{6}(\text{J}/\text{m}^3)$ and $D=3.0\cdot10^{-3}(\text{J}/\text{m}^2)$. The competition of these energies is responsible for the magnetic configurations observed in the nanodot. In order to obtain skyrmion clusters, we place in an initially saturated configuration in the $+z$-direction, $N$ circular and localized antiparallel regions (in the $-z$-direction). This magnetic configuration is relaxed by minimizing the total energy $U[\bm m]$ until the condition $\bm{m}\times \bm{H}_{\text{eff}}=0$ is satisfied and a steady state is reached (see Fig. \ref{fig1} b). The skyrmions tend to be organized in symmetrical clusters (with cluster radius $R$) and the radius of their cores ($r$) do not vary much, thus avoiding the deformation of the skyrmions and the excitations of internal modes of oscillations that, in fact, have been studied in various contexts \cite{PhysRevLett.108.017601, LIU20189, doi:10.1063/1.5010948}. It is worth noting that, in this study, we have considered the core radius of the skyrmions (from now on, skyrmion radius) as the length along the $x-y$ plane in which $m_z$ continuously varies from -1 to 0. Finally, we have discretized the system into cubic cells of 1 nm$^3$.

\section*{Initial configuration disturbed by a spin-polarized current}

A spin-polarized current interacts with the magnetization of the sample by coupling the spin of the itinerant electrons that the current carries and the localized electrons that produce the magnetization \cite{Slonczewski-1996lq}. As a result of the conservation of angular momentum, a torque is produced between the current and the magnetization called spin-transfer torque. In this work we are interested in the effect produced by a perpendicular polarized current and we consider a homogeneous magnetization along the z-axis. As a result, there is a torque only when the direction of the magnetization is non-collinear with the spin polarization. In other words, we neglect the dynamics of magnetization along the thickness of the layer. If we define a particular direction for the polarization of the perpendicular current, $\hat{\bm  p}$, the torque takes the form of the so-called Slonczewski torque \cite{Slonczewski-1996lq}:
\begin{equation}
 \bm{\Gamma}_{\text{ST}}=\gamma  \sigma J \varepsilon \bm{m} \times \left( \bm{m}\times \hat{\bm  p}  \right),
 \label{eq_ST}
\end{equation}
where $J$ is the magnitude of the current density, $\sigma=\hbar/(\mu_{0}e M_s\Delta)$ with $\hbar$ the Planck's constant reduced, $e$ is the magnitude of the electron charge and $\Delta$ is the thickness of the magnetic layer. The definition of $\varepsilon$ is more complicated, including transport coefficients \cite{PhysRevB.70.172405}. In the standard approach, this depends explicitly on the degree of polarization $P$ of the current, the conductance and the effective resistances of the layers \cite{PhysRevB.70.172405}. 

Additionally, we introduce the field-like torque, which is a secondary torque that appears when a vertical current is injected into the system \cite{PhysRevB.71.024411, PhysRevLett.102.067206}:
\begin{equation}
\bm{\Gamma}_{\text{FLT}}=- \gamma  \sigma J  \varepsilon'  (\bm{m} \times \bm{p}),
\label{eq_FLT}
\end{equation}
where the value of $\varepsilon'$, commonly called secondary spin transfer term, is not yet clearly defined \cite{chanthbouala-2011aa} and, thus, we study its influence considering different values. The action of field-like torque is, as the name implies, functionally equivalent to the effect of an external magnetic field. This torque allows to move a domain wall with a current two orders of magnitude lower than that necessary in the absence of this torque, which captures its interest in consideration in future applications \cite{chanthbouala-2011aa}. Although many studies have investigated the influence of spin-polarized currents on skyrmions, none of them have explored the influence of field-like torque, an issue that may be fundamental in the development of future spintronic devices.

\begin{figure}[h]
\centering
\includegraphics[width=.5\textwidth]{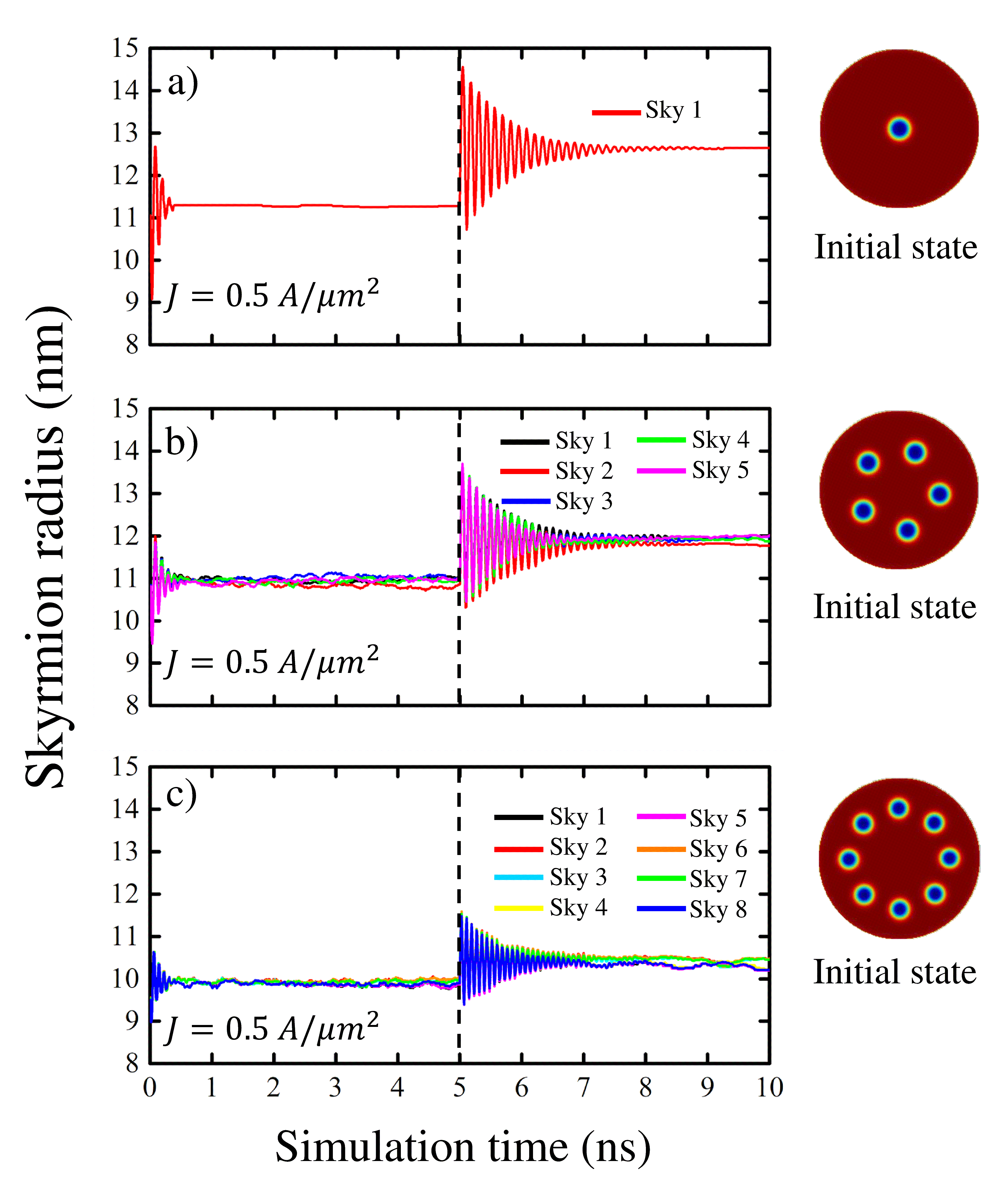}
\caption{(Color online) Temporal evolution of the skyrmion radius for three different configurations investigated when a spin-polarized current is applied perpendicular to the nanodot for 5 ns. The inserted snapshots represent the initial magnetic state before applying the current pulse. Once the current is turned off, the magnetic states relax to their equilibrium positions.}
\label{fig2}
\end{figure}

Once the system has relaxed to the desired magnetic configuration (with a certain number $N$ of skyrmions), a spin-polarized current perpendicular to the nanodot plane and homogeneously polarized with $ \bm{ \hat p}=+\bm{\hat z}$ is introduced during 5 ns. In this work, we use a standard value of $P=0.4$, obtaining an $\varepsilon=0.2$ \cite{Thiaville-2005cr}. In a first series of simulations, we consider a current density of $J=0.5$ A/$\mu$m$^2$, whose order of magnitude is a standard value used in experimental works \cite{doi:10.1063/1.4930904, Zhang2015}, and a secondary spin transfer term given by $\varepsilon'=0.02$, which has been reported in \cite{PhysRevLett.102.067206, sampaio-2013aa}. This current produces a decrease in the skyrmions radius due to the effect of $\text{FLT}$ (see left section of Fig. \ref{fig2}).

\newpage
\section*{Relaxation process after current is turned off: oscillations}

Once the current pulse is turned off, the system relaxes until it reaches its equilibrium state again after approximately
3 ns (see Fig. \ref{fig2}). The relaxation time is related to dissipative dynamics, mainly controlled by the $\alpha$ parameter. From Fig. \ref{fig2} we can see that for all the cases considered (1, 5 and 8 skyrmions) the relaxation process is mediated by oscillations of the skyrmion sizes, which are in phase, and follow a behavior similar to damped harmonic oscillator (for more details, see Fig. S1 of the Supplementary Information). Although skyrmions are known to exhibit several internal modes, typically, the lowest frequency local mode is the breathing mode \cite{PhysRevB.90.064410}, in which skyrmion size is oscillating around it's equilibrium value. From analysis of Fig. \ref{fig2} we conclude that the observed oscillations correspond to the breathing mode. It is important to note that, once the current pulse is turned off, the skyrmions tend to relax to their initial size.

\subsection*{Effect of field-like torque on the dynamics of skyrmions}

\begin{figure}[h!]
\centering
\includegraphics[width=.5\textwidth]{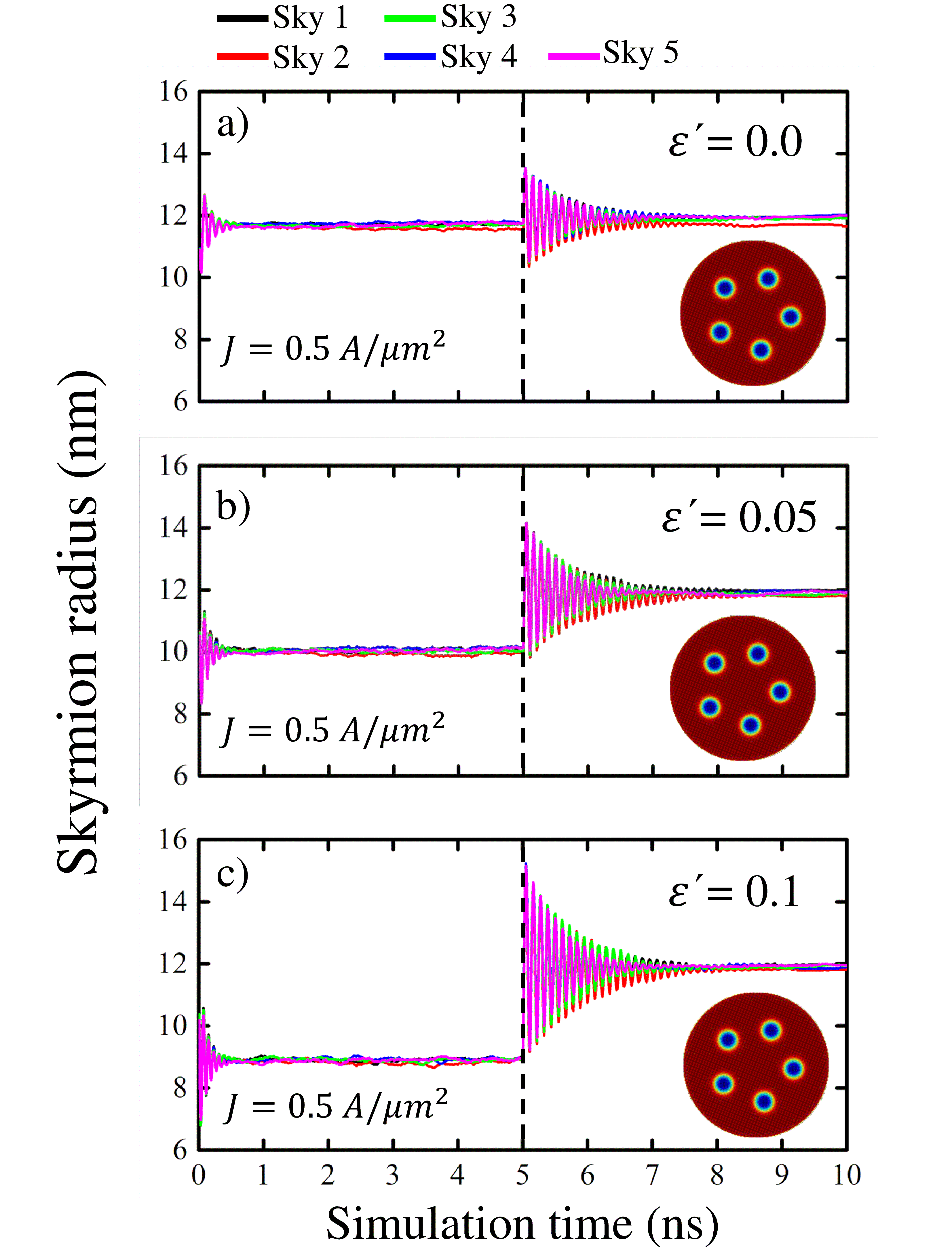}
\caption{(Color online) Temporal evolution of the skyrmion radius of a cluster of 5 skyrmions when a spin-polarized current is applied perpendicular to the nanodot for 5 ns with a) $\varepsilon'=0.0$, b) $\varepsilon'=0.05$ and c) $\varepsilon'=0.1$. The inserted snapshots represent the initial magnetic state before applying the current pulse. Once the current is turned off, the magnetic state relaxes to its equilibrium position.}
\label{fig3}
\end{figure}

To understand the effect of field-like torque on the skyrmions dynamics, in Fig. \ref{fig3} we show the behavior of the skyrmions for three values of $\varepsilon'$. We can see that as the field-like torque increases, the skyrmion radius decreases because the field produced by $\text{FLT}$ is opposite to the skyrmion core (in the presence of a spin-polarized current). This produces that when the current is turned off, the skyrmions that had drastically reduced their size due to the field-like torque now relax oscillating with a greater amplitude.

\begin{figure}
\centering
\includegraphics[width=.55\textwidth]{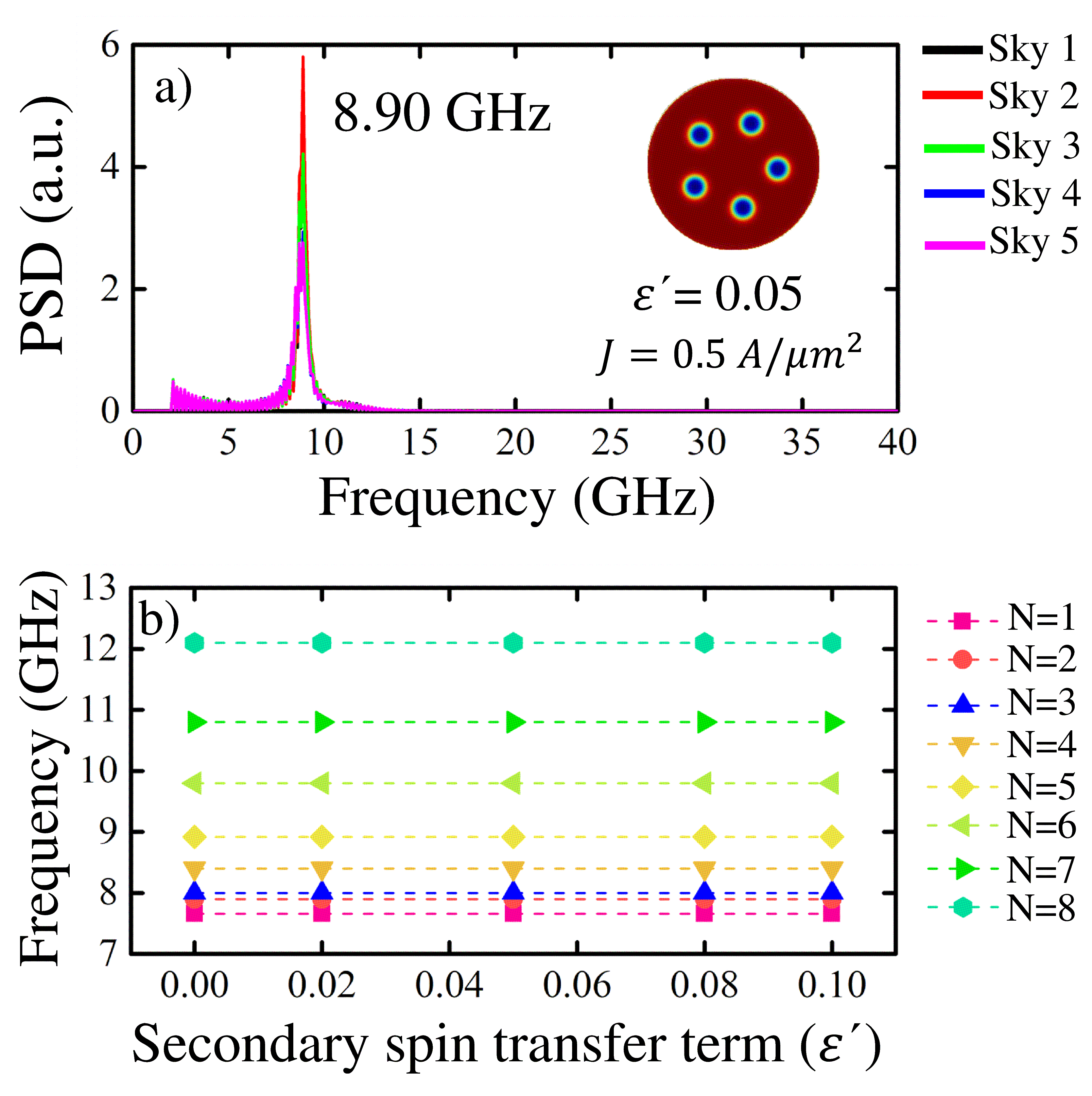}
\caption{(Color online) a) Power spectral density (PSD) for a cluster of 5 skyrmions when a spin-polarized current is applied with $\varepsilon'=0.05$. b) Evolution of the response frequencies as a function of the secondary spin transfer term $\varepsilon'$ after the system is excited by a spin-polarized current applied perpendicular to the nanodot, for $N$ skyrmions hosted in a multilayer nanodot. For each value of $N$, the skyrmions oscillate with the same frequency.} 
\label{fig4}
\end{figure}

The oscillation frequencies of the skyrmions can be analyzed by calculating the power spectral density (PSD) of each signal associated with each skyrmion. Fig. \ref{fig4} shows the response frequency for different values of $N$. In Fig. \ref{fig4}a, we  show the PSD of the oscillations of 5 skyrmions with $\varepsilon'=0.05$ analyzed in Fig. \ref{fig3}b. Our results show that the oscillation frequencies of each skyrmion are positioned at a very precise point in the frequency space ($8.90$ GHz). Our calculations also show that this phenomenon repeats for $\varepsilon'=0.0$ and $\varepsilon'=0.1$ (see Fig. S2 of the Supplementary Information). On the other hand, Fig. \ref{fig4}b presents the peak of the oscillation frequency of the skyrmions considering different values of $\varepsilon'$. It is possible to note that the effect of $\varepsilon'$ has no influence on the natural frequency of the system and is limited exclusively to a change in the initial amplitude of the oscillation, which reveals that the relevant parameter here is the Slonczewski torque.

It is worth noting that the coherence in the oscillations occurs mainly due to the small size of the skyrmions and because they interact weakly with each other (for more details, see Fig. S3 of the Supplementary Information). The fact that there is a specific frequency of oscillation suggests that it is associated with the total energy of the system and not with the energy of a specific skyrmion. In this way, we can obtain different oscillation frequencies for the same nanodot simply by changing the number of skyrmions present without modifying any geometrical or magnetic parameter. This advantage may be of practical interest for potential applications in magnonics and spintronics.

\begin{figure}[h]
\centering
\includegraphics[width=.8\textwidth]{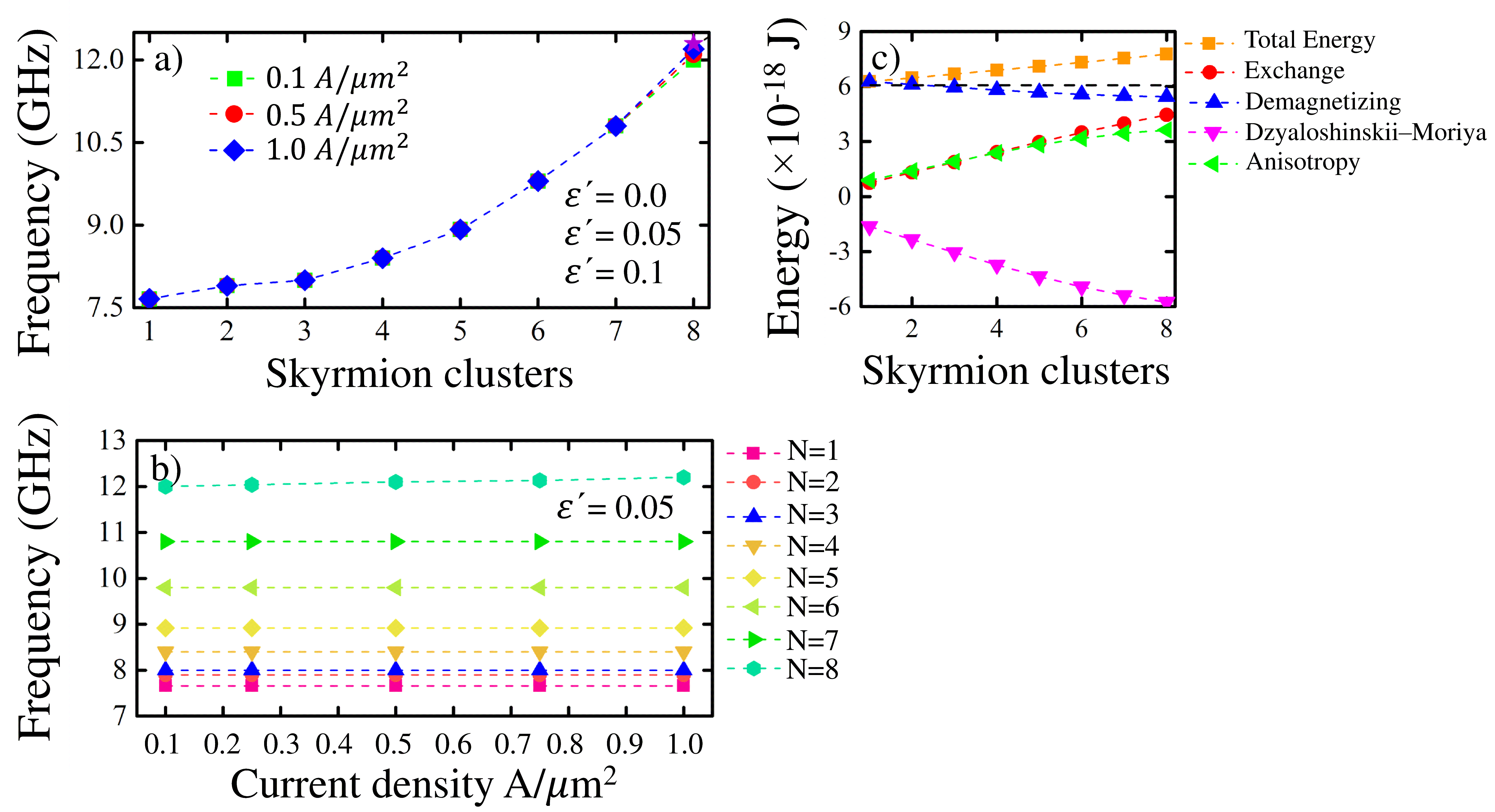}
\caption{(Color online) Evolution of the response frequencies as a function of a) the number of skyrmions $N$ hosted in a multilayer nanodot and b) the applied current density $J$, after the system is excited by a spin-polarized current applied perpendicular to the nanodot. In a) three values of $J$ and three values of $\varepsilon'$ were considered, while in b) we use $\varepsilon'$=0.05. c) Magnetic energies as a function of $N$. The segmented horizontal line without geometric symbols corresponds to the uniform ferromagnetic configuration in the $+z$-direction.}
\label{fig5}
\end{figure}

Our results are summarized in Fig. \ref{fig5}a which shows the peak frequency for three different values of $\varepsilon'$ and for three different values of the current density $J$ as a function of the number of skyrmions. As we can see, the value of $\varepsilon'$ is irrelevant (the curves overlap) for any value of $N$, while the value of $J$ begins to be relevant only from $N = 8$, obtaining that a higher current density produces a greater oscillation frequency. The above result is due to the fact that the field-like torque also depends on the value of $J$ (see Eq. (\ref{eq_FLT})). If the action of the field-like torque is intense, the oscillations amplitude going to be greater when the current pulse is turned off, becoming highly interacting system due to the increase in skyrmion sizes. The same situation occurs for large values of $\varepsilon'$. Nevertheless, this interaction can be reduced by controlling the skyrmion sizes through homogeneous magnetic fields, as studied in reference \cite{tejo2018}. Additionally, we can confirm our results from Fig. \ref{fig5}b in which it is possible to observe that, for a fixed value of $\varepsilon'$, the oscillation frequency only depends on $N$, except for $N = 8$ in which we found a very small linear dependency with $J$. This result prevents us from knowing the possible oscillations that appear in a cluster of more numerous skyrmions, since the oscillation would depend on $J$. In addition, it is important to note that for $N \ge 9$ the array symmetry is lost and the interaction between the skyrmions becomes important, causing that the radius of the skyrmions that form the cluster are not equal (see Fig. S4 of the Supplementary Information).

From a theoretical point of view, we can understand the behavior of the breathing mode (its natural frequency of oscillation) as a consequence of the change in radius of the skyrmion produced by the number of these. In fact, skyrmions repel each other \cite{Capic2020} producing symmetrical patterns when many of them are forced to be in the same nanodot, causing a reduction in the radius of their cores. Additionally, it is known that the frequency of the uniform breathing mode exhibits linear growth with respect to a magnetic field antiparallel to the skyrmion core direction \cite{PhysRevB.90.064410, Lin2014}. This result can also be interpreted as a direct dependence of the frequency of the uniform breathing on the radius of the skyrmion cores. Fig. S5 of the Supplementary Information allows to verify the linear relationship that exists between the radius of the core of the samples considering a different number $N$ of skyrmions and the frequency of the oscillations.

Finally, it is also interesting to explore the relation between the frequency and the total energy of the system. Fig.\ref{fig5} c shows the energies of the system as a function of the number of skyrmions. From this figure we can see how the DMI and demagnetizing energy decrease almost linearly with the increase in the number of skyrmions, while the exchange and anisotropy energies follow the opposite behavior. The sum of all these energy contributions leads us to the fact that the total energy of the system increases as the number of skyrmions in the system increases, a behavior similar to that observed for the system's response frequency, which justifies the relationship between the response frequency and total energy that we have proposed throughout this article.

\newpage
\section*{Conclusions and perspectives}

In conclusion, we have investigated the behavior of skyrmion clusters under a controlled situation, where the interaction between skyrmions is weak. For this, we have chosen the parameters associated with a Pt/Co multilayer nanodot and introduced an initial condition characterized by a multiskyrmion solution. By relaxing these initial conditions, we have been able to achieve stable configurations of skyrmion clusters that can be stimulated by a spin-polarized current applied perpendicular to the nanodot, which interacts with the magnetic layer through the Slonczewski torque and the field-like torque. We found that, once the current pulse is turned off, the relaxation to the equilibrium state occurs through coherent oscillations (all skyrmions oscillate in phase with the same frequency) that are exclusively related to the number $N$ of skyrmions in the weakly interacting regime. The fact that skyrmions interact weakly with each other is crucial to obtain coherent oscillations. We have observed that in other cases, where the interaction is not negligible ($N \ge 9$), the oscillations become more complex and cannot be directly related to the number $N$ of skyrmions in the nanodot. However, these new effects will be explored in future work, where we will surely obtain rich phenomena due to the superposition and interference of the emitted spin waves. 

Finally, it could be interesting to explore the relation between the response frequency and the possible appearance of a response frequency by the injection of spin waves of different frequency. It is natural to suppose that this two frequencies must coincide but at this point we are not able to assure that.

\section*{Acknowledgments}

The authors acknowledge financial support from Dicyt-Usach 041931EM-POSTDOC, Basal Project  AFB180001, and Fondecyt 11171122 and 1200302.

\section*{Author contributions}
The problem was designed and proposed by Felipe Tejo and Juan Escrig. The micromagnetic simulations were developed by Felipe Tejo. The algorithm that allows the analysis of the results was developed by Felipe Velozo. All the authors analysed the results. The article was written by Ricardo Gabriel Elías and Felipe Tejo. Finally, the article was reviewed by all authors.

\end{document}